\documentclass{PoS}

\title{The domain of validity of fluid dynamics and the onset of cavitation
in ultrarelativistic heavy ion collisions}

\ShortTitle{Onset of cavitation
in ultrarelativistic heavy ion collisions}

\author{\speaker{Gabriel S.~Denicol}\\
        Department of Physics, McGill University, 3600 University Street, Montreal,
QC, H3A 2T8, Canada \\
        E-mail: \email{denicol@physics.mcgill.ca}}

\author{Charles Gale\\
        Department of Physics, McGill University, 3600 University Street, Montreal,
QC, H3A 2T8, Canada }

\author{Sangyong Jeon\\
        Department of Physics, McGill University, 3600 University Street, Montreal,
QC, H3A 2T8, Canada }

\abstract{If the bulk viscosity of QCD matter is large, the effective pressure of the hot and dense matter created in ultrarelativistic heavy ion collisions can become
negative, leading to instabilities in the evolution of the plasma. In the context of heavy ion collisions, this effect is sometimes referred to as cavitation. 
In this contribution we discuss the onset of cavitation in event-by-event hydrodynamic simulations of ultrarelativistic heavy ion collisions at LHC energies. We estimate how large the bulk viscosity of the QGP has to be in the QCD (pseudo) phase transition region in order for the effective pressure of the system to actually become negative.}

\FullConference{9th International Workshop on Critical Point and Onset of Deconfinement - CPOD2014,\\
		17-21 November 2014\\
		ZiF (Center of Interdisciplinary Research), University of Bielefeld, Germany}

\begin{document}

\section{Introduction}
Ultrarelativistic heavy ion collisions at the
Relativistic Heavy Ion Collider (RHIC) and the Large Hadron Collider (LHC)
are able to reach temperatures high enough to create and study the
quark-gluon plasma (QGP) in a controlled experimental environment. So far,
fluid-dynamical descriptions that take into account only the effects of shear
viscosity have been considered sufficient to describe the time evolution of
the QGP created in such collisions.

However, QCD is a nonconformal field theory and, in principle, one should
not neglect the effects of bulk viscosity on the dynamics of the QGP. The
bulk viscosity is expected to be maximum near the (pseudo) QCD phase
transition and, if it is sufficiently large, it can drive the effective pressure of the system to become negative
\cite{Torrieri:2008ip,Rajagopal:2009yw}, driving fluid dynamics out of its domain of validity and changing the
freeze-out process of the system. This effect was referred to as cavitation in Ref.~\cite{Rajagopal:2009yw}.

In this contribution we discuss the onset of cavitation in event-by-event
hydrodynamic simulations of heavy ion collisions at LHC energies. We find that if the bulk viscosity coefficient
becomes large even in a small temperature domain, $170<T<190$ MeV, it can already considerably affect the evolution of the system.
We further estimate the value of bulk viscosity, around this narrow temperature region, that is required to make the effective pressure
become negative.   

\section{Model}

The time evolution of the hot QCD matter produced is solved
using relativistic dissipative fluid dynamics. The main equations of
motion are the continuity equation for the energy-momentum tensor, $T^{\mu
\nu }$, $\partial _{\mu }T^{\mu \nu }=0$, where $T^{\mu \nu }=\varepsilon
u^{\mu }u^{\nu }-\Delta ^{\mu \nu }(P_{0}+\Pi )+\pi ^{\mu \nu }$, with $%
\varepsilon $ being the energy density, $P_{0}$ the thermodynamic pressure, $u^{\mu }$ the fluid velocity, $%
\Pi $ the bulk viscous pressure, and $\pi ^{\mu \nu }$ the shear-stress
tensor. Here, we introduced the projection operator $\Delta ^{\mu \nu }=g^{\mu \nu
}-u^{\mu }u^{\nu }$ onto the 3-space orthogonal to the fluid velocity. 
The equation of state, $P_{0}(\varepsilon )$, reflects chemical equilibrium and is taken from Ref.~\cite{Huovinen:2009yb}, corresponding to a parametrization of a lattice QCD calculation matched to a hadron resonance gas calculation at lower
temperatures. We assume that the baryon number density and diffusion are zero at all
space-time points and our metric convention is $g^{\mu \nu }=\mathrm{diag}(+1,-1-1-1)$.

The time-evolution equations satisfied by $\Pi $ and $\pi ^{\mu \nu }$ are derived from kinetic theory \cite{Denicol:2012cn,Denicol:2014vaa} and solved numerically using the \textsc{music} hydrodynamic simulation \cite{MUSIC,Marrochio:2013wla}. We solve
\begin{equation}
\tau_{\Pi }\dot{\Pi}+\Pi = -\zeta \theta -\delta _{\Pi \Pi }\Pi \theta
+\lambda _{\Pi \pi }\pi ^{\mu \nu }\sigma _{\mu \nu }\;, \label{intro_1}
\end{equation}
\begin{equation}
\tau_{\pi }\dot{\pi}^{\left\langle \mu \nu \right\rangle }+\pi ^{\mu \nu } = 2\eta \sigma ^{\mu \nu }-\delta _{\pi \pi }\pi ^{\mu \nu }\theta +\varphi
_{7}\pi _{\alpha }^{\left\langle \mu \right. }\pi ^{\left. \nu \right\rangle
\alpha } 
-\tau _{\pi \pi }\pi _{\alpha }^{\left\langle \mu \right. }\sigma
^{\left. \nu \right\rangle \alpha }+\lambda _{\pi \Pi }\Pi \sigma ^{\mu \nu
}.  \label{intro_2}
\end{equation}

For the sake of simplicity, most of our transport coefficients are calculated
using the Boltzmann equation near the conformal limit \cite{Denicol:2014vaa}.
For the shear viscosity, we use the temperature dependent coefficient (HQ-HH parametrization) defined in Ref.~\cite{Niemi:2012ry}.
For the bulk viscosity, we consider two parametrizations: the bulk viscosity defined
in Refs.~\cite{Denicol:2009am,Ryu:2015vwa}, which corresponds to a parametrization
of calculations from \cite{Karsch:2007jc}, for the QGP phase, and \cite{NoronhaHostler:2008ju} for the hadronic phase.
These two calculations are matched around $T_c=180$ MeV and the value of $\zeta /s$
at this temperature is $\zeta /s(T_c)\approx 0.3$. The other choice of bulk viscosity is a parametrization
of calculations from lattice QCD \cite{Rajagopal:2009yw}, for the QGP phase, and, once more, \cite{NoronhaHostler:2008ju} for the hadronic phase.
In this case, the calculations are matched around $T_c=180$ MeV with a much larger value of $\zeta /s$, $\zeta /s(T_c)\approx 1$. So both
parametrizations of $\zeta /s$ are the same in the hadronic phase and are very similar in the QGP phase (at high temperatures, both choices of $\zeta /s$ 
are basically zero). However, they differ considerably in the phase transition region, for temperatures $170<T<190$ MeV. In the following we will see that, 
even though the viscosities are only different in such narrow temperature domains, they are still able to considerably modify the time-evolution of the system.
Both parametrizations are shown in Fig.~1(a). 

\begin{figure*}[th]
\includegraphics[width=8cm]{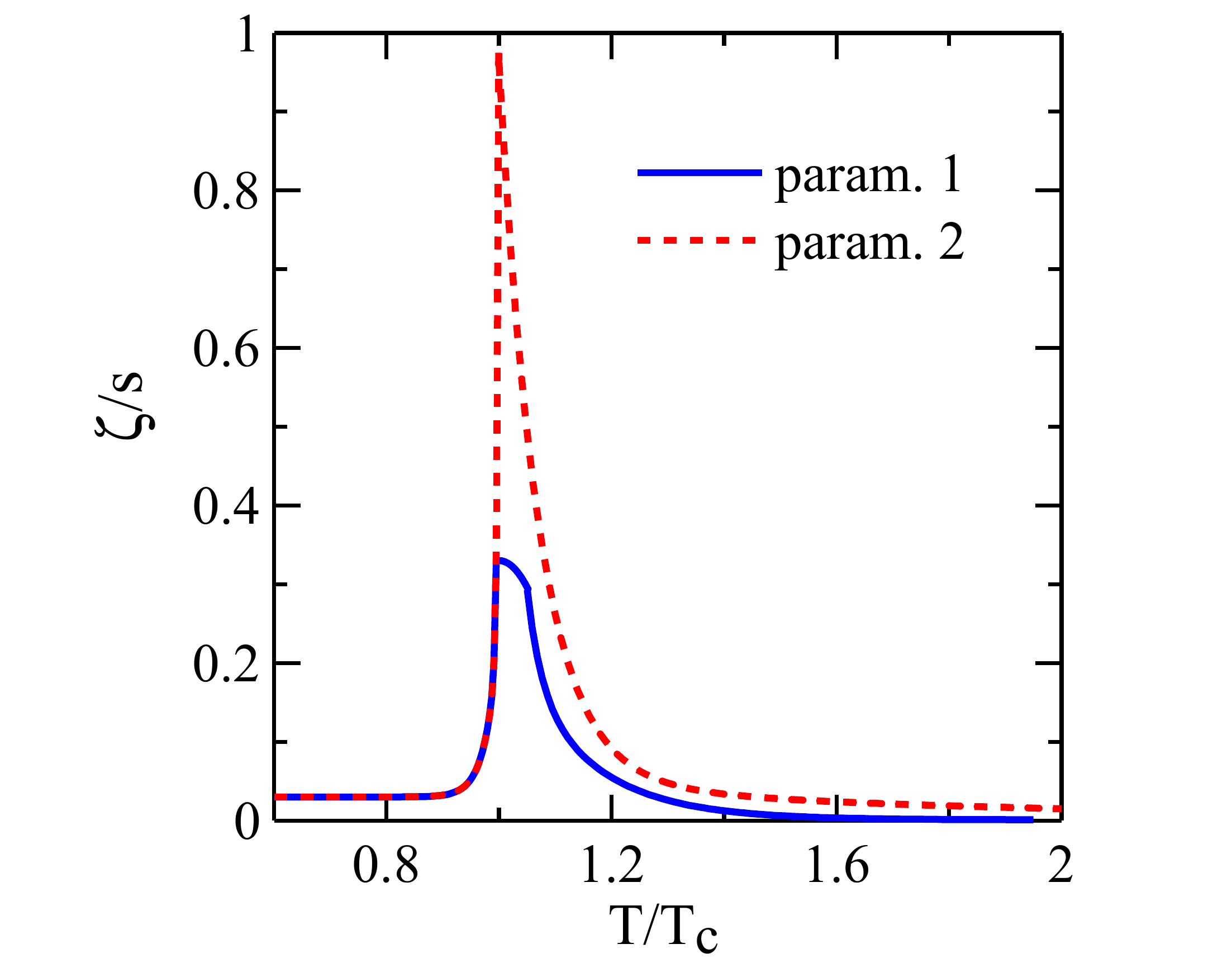}
\includegraphics[width=8cm]{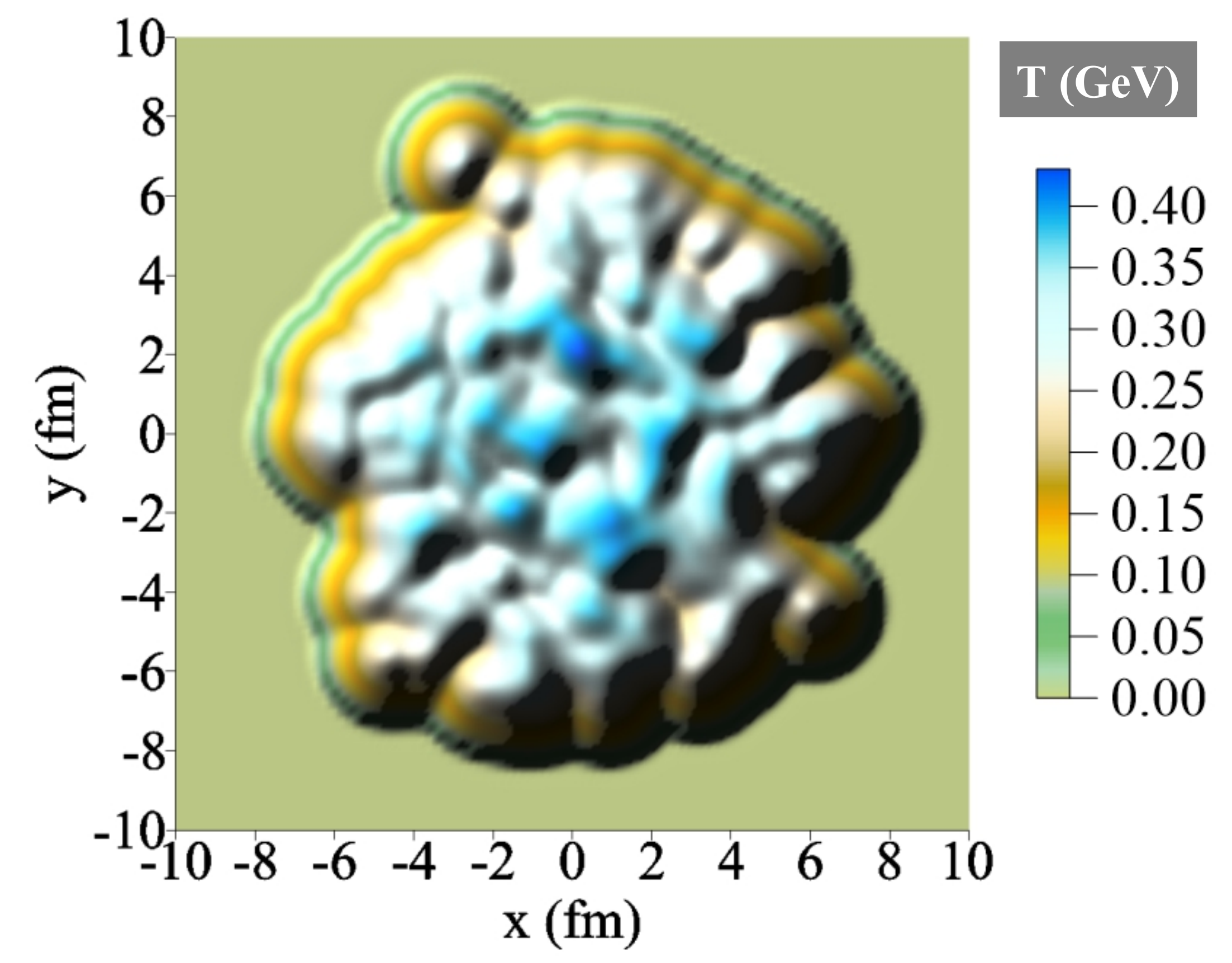}
\caption{(Color online) The left panel shows the bulk viscosity over entropy density ratio as a function of temperature (normalized by $T_c=180$ MeV). The solid blue line
corresponds to parametrization 1 and the dashed red line to parametrization 2. The right panel shows the initial temperature profile employed in all simulations
discussed in this work.}
\label{fig1}
\end{figure*}

As a test case, we consider the hydrodynamic simulation of one perfectly central (impact parameter equal to zero) Pb-Pb
collision event at LHC energies. The initial state considered is calculated
using the MC-Glauber model, with the initial energy density of the system being proportional to the density of participants. The initial
time is taken to be $\tau _{0}=0.6$ fm and we further assume that the system
starts at rest and in local thermodynamic equilibrium, i.e., $u^{i }\left(
\tau _{0}\right) =\Pi \left( \tau _{0}\right) =\pi ^{\mu \nu }\left( \tau
_{0}\right) =0$. The initial entropy of the medium is determined in such a
way that the multiplicity of charged hadrons measured at the LHC is described. The initial temperature profile is plotted in Fig.~1(b).

\begin{figure*}[tbp]
\includegraphics[width=14cm]{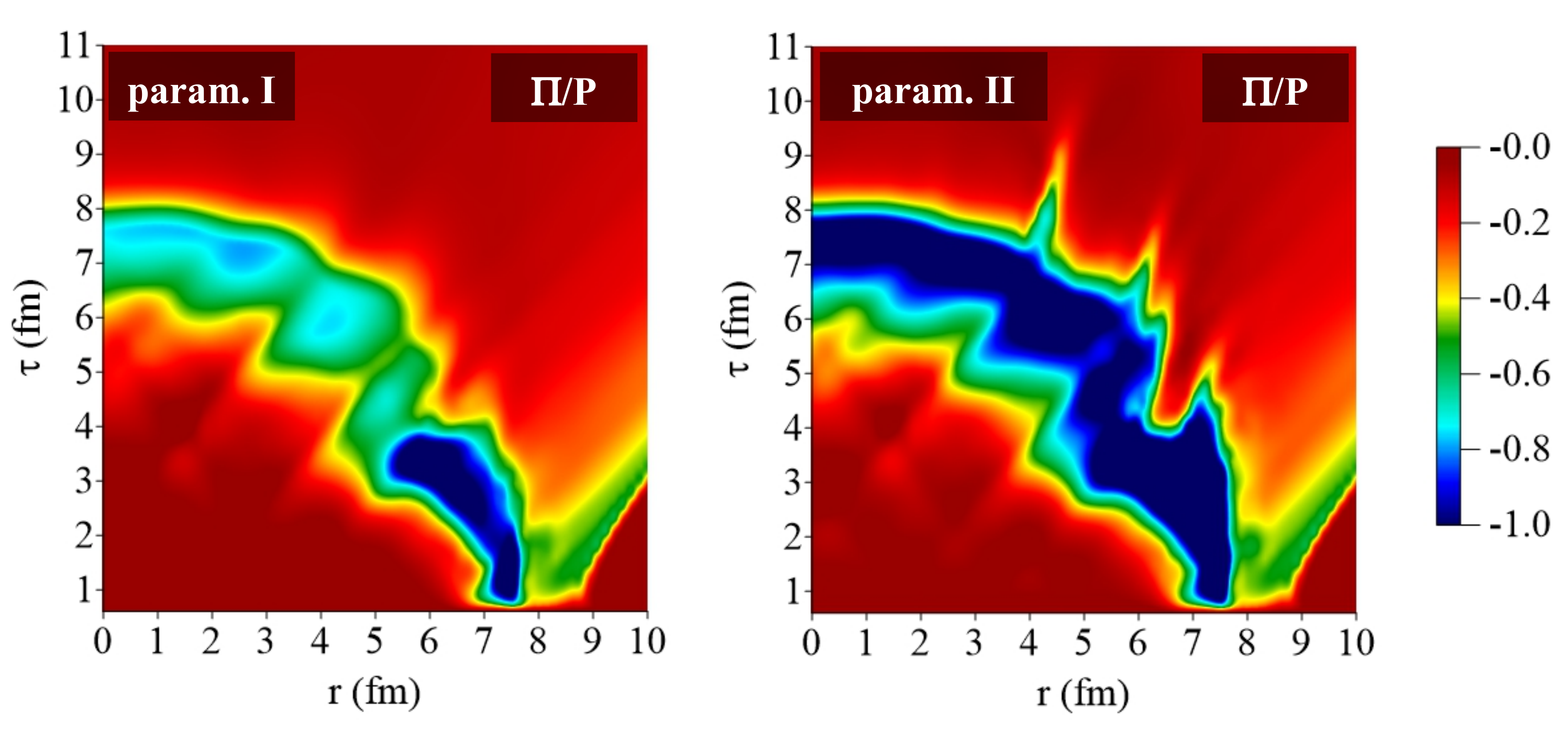}
\caption{(Color online) Spacetime evolution of $\Pi/P_{0}$ in a hydrodynamic simulation of a central Pb-Pb collision. The left panel shows the 
simulation that employed parametrization 1 of $\zeta/s$ while the right panel shows the same simulation using parametrization 2.}
\label{fig:2}
\end{figure*}

\begin{figure*}[tbp]
\includegraphics[width=16cm]{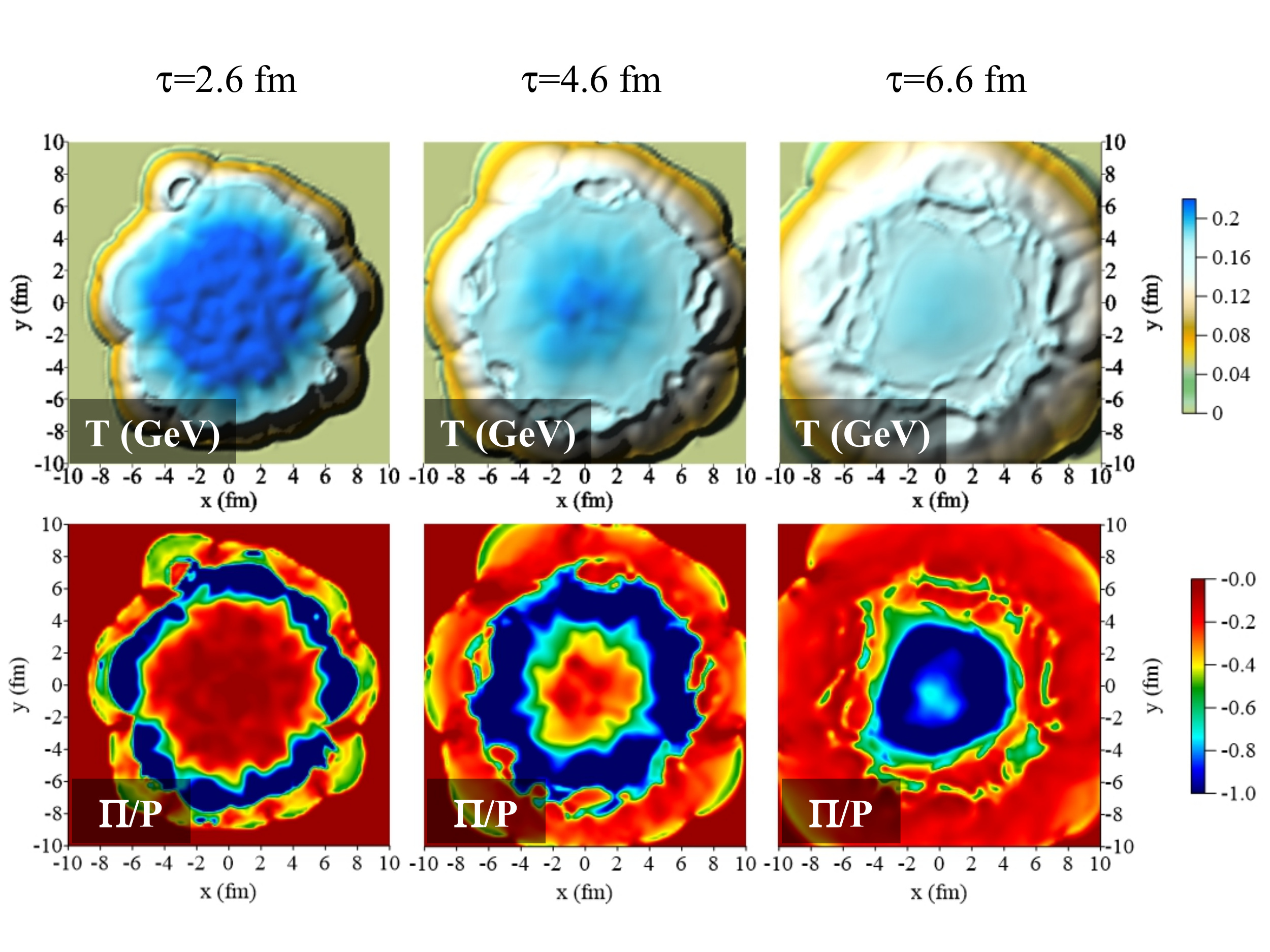}
\caption{(Color online) Spatial profiles of temperature (upper panels) and $\Pi/P_{0}$ (lower panels) for three time steps of our simulation, 
$\tau=2.6$ fm (left panels), $\tau=4.6$ fm (middle panels), and $\tau=6.6$ fm (right panels). }
\label{fig:3}
\end{figure*}

\section{Results}

In Figs.~2(a) and (b) we show the space-time profile of $\Pi /P_{0}$ in the
chosen hydrodynamic event for the two parametrizations of $\zeta /s$
shown in Fig.~1. The $\Pi /P_{0}$ profiles are shown in the $\left( \tau ,r=\sqrt{x^{2}+y^{2}%
}\right) $ plane, along the $x=y$ axis.

We see that the effective pressure of the system, $P_{\rm{eff}} = P_0 + \Pi$, can become very low in some spacetime points for both parametrizations of $\zeta /s$ considered.
Such small values of pressure occur exactly in the phase transition region, where the bulk viscosity has a peak. The second parametrization, which has the largest bulk viscosity around the phase transition region, leads to the formation of a negative pressure
hypersurface that encloses the whole system. That is, it is impossible for a
fluid element to reach the constant temperature freezeout hypersurface (for
all freezeout temperatures, $T_{\rm{fo}}<180$ MeV) without passing through a negative pressure region. However,
for parametrization 1, where $\zeta /s<0.3$, even though the effective pressure can reach small values, it is still positive in most space-time regions. We have checked that, by increasing the value of bulk viscosity around $T_c$ in parametrization 1 from $\zeta /s(T_c)=0.3$ to $\zeta /s(T_c)=0.6$, cavitation starts to occur.

In Figs.~3 we show the transverse spatial profile of the temperature and bulk viscous pressure over thermodynamic pressure
for three different time steps, $\tau=2.6$ fm (left panels), $\tau=4.6$ fm (middle panels), and $\tau=6.6$ fm (right panels). The upper panels show the temperature profiles while the lower panels display the $\Pi /P_{0}$ profiles. This time evolution was calculated with the second parametrization of $\zeta/s$, which has the largest
value around $T_c$, $\zeta/s(T_c) \approx 1$. We already showed in Fig.~2 that regions with negative effective pressure will occur in the fluid when such large values of bulk viscosity are employed. 

We see that, at the initial times, there is always a ring of negative effective pressure formed near the edges of the system. These rings of negative pressure gradually propagate towards the center of the fluid and, at $\tau=6.6$, the center of the fireball is already filled by a negative pressure domain. In the temperature profiles shown in the upper panels of Fig.~3, we see the formation of "holes" or pockets of small temperatures created in the system due to the negative effective pressure achieved in these regions. Also, we see that the bulk viscosity is so large that it considerably reduces the expansion rate of the system, leading to a very uniform temperature profile in the center of the fireball. These effects resembles what happens in a first order phase transition, where the expansion of the system considerably slows down in the mixed phase region.     

\section{Conclusions}
In this work, we investigated the onset of cavitation in event-by-event hydrodynamic simulations of heavy ion collisions. We showed that, if the bulk viscosity is
of the order of $\zeta/s \approx 1$ around the QCD phase transition region, the fluid-dynamical evolution of the system will generate negative effective pressures
in certain regions of the fluid. This behavior indicates the breakdown of the fluid-dynamical description and, in the context of hydrodynamic simulations of heavy ion collisions, will lead to changes in the way the freezeout process of the system is developed. We have also found that, if the bulk viscosity is $\zeta/s \approx 0.3$ around the QCD phase transition region, cavitation will not happen, but the bulk viscosity will have significant effects on the evolution of the system.

This work was supported by the Natural Sciences and Engineering
Research Council of Canada. G.~S.~Denicol acknowledges support through a Banting
Fellowship of the Natural Sciences and Engineering Research Council of Canada.

\end{document}